\documentclass[a4paper]{jpconf}
\usepackage{graphicx}
\begin{document}
\title{Chemistry in AGB stars: successes and challenges}

\author{T J Millar}

\address{Astrophysics Research Centre, School of Mathematics and Physics, Queen's University 
Belfast, University Road, Belfast BT7 1NN, UK}

\ead{tom.millar@qub.ac.uk}

\begin{abstract}
Emission and absorption line observations of molecules in late-type stars are a vital component 
in our understanding of stellar evolution, dust formation and mass loss in these objects.  
The molecular composition of the gas in the circumstellar envelopes of AGB stars reflects 
chemical processes in gas whose properties are strong functions of radius with density and 
temperature varying by more than ten and two orders of magnitude, respectively.  In addition, 
the interstellar UV field plays a critical role in determining not only molecular abundances 
but also their radial distributions. In this article, I shall briefly review some recent 
successful approaches to describing chemistry in both the inner and outer envelopes and 
outline areas of challenge for the future.
\end{abstract}

\section{Introduction} \label{intro}
The circumstellar envelopes (CSEs) of AGB stars have long been known to present a rich molecular 
chemistry dominated by the interaction of external, interstellar FUV photons with parent 
species formed by thermal equilibrium processes near the photosphere \cite{hug82, nej84,
hug84, nej87, che93a, che93b}.  These, and more recent models \cite {mil00, agu08, 
cor09, agu10}, have included more accurate descriptions of the physical conditions 
through the inclusion of clumps and density-enhanced rings in the CSE around the carbon-rich 
AGB star IRC+10216 (CW Leo) and through the addition of an extensive chemistry to describe the
anions recently detected therein. The result is a consensus that the chemistry of the 
external envelope of IRC+10216, and by extension all AGB CSEs, is a photon-dominated 
process, a process whose final molecular products give information on mass-loss history, 
wind acceleration, dust formation, dredge-up and nucleosynthesis.

As well as studies of the chemical processes in the outer CSE, there have also been 
investigations of the the interaction between physics and chemistry in the inner CSE. For example, 
pioneering work on the chemistry induced by shock waves driven by stellar pulsations, 
\cite{che92, dua99}, has been extended \cite{che12, gob16} to include the formation of 
new species and dust grain formation. These papers show that if shocks are strong then any
molecules formed in thermodynamic equilibrium (TE) are rapidly destroyed in the immediate 
post-shock gas and that `parent' species available for chemistry in the outer CSE are the 
end products of shock chemistry coupled with dust nucleation and growth.
Challenges in understanding astrophysics and astrochemistry are, as ever in astronomy, driven 
by advances in observational techniques, instruments and facilities, most recently from
the {\it Herschel Space Observatory} and {\it ALMA}.

\section{Circumstellar chemistry} \label{cschem}
In this section I will briefly review some progress made in the chemistries of both O-rich
and C-rich AGB stars and give some indication of current challenges.

\subsection{O-rich late-type stars} \label{orich}
{\it Herschel} and {\it ALMA} have both given us remarkable new information on the chemistry of O-rich CSEs,
especially in the internal layers close to the photosphere. A large number of detailed observational studies 
\cite{dec10a, kim10, ten10a, ten10b, deb13} have both increased the range of 
species detected near the photosphere and provided much improved abundance estimates. These
include studies of the refractory species SiO, TiO, TiO$_2$, AlO and AlOH, thought to be 
involved in the creation of silicate grains in these stars. Gobrecht et al. 
\cite{gob16} have presented a very detailed model of dust formation in the O-rich star IK Tau.
They considered chemical reactions in the shocked gas created by periodic stellar pulsations
including the formation of small magnesium silicate and alumina clusters. Their best-fit
model to observations starts with a periodic shock wave at a velocity of 32 km s$^{-1}$
at a radius of 1 R$_{\star}$ which propagates outward with a velocity proportional to
$r^{-2}$, where $r$ is the radial distance from the photosphere. For radii less than 2 R$_{\star}$
the immediate post-shock densities are very high, greater than 10$^{13}$ cm$^{-3}$, and
solidly in the regime where three-body reactions must be considered. The immediate post-shock 
temperature is also very high, more than 4000 K. At these densities and temperatures any
molecules formed at thermal equilibrium (TE) in the photosphere are destroyed. 
Over a pulsational phase the density and temperature both fall and new molecules, whose 
compositions and abundances depend on chemical kinetics in a cooling, expanding flow, form.
In this post-shock gas the chemistry is dominated by high temperature neutral-neutral reactions.

It is, of course, not surprising that in O-rich  AGB stars, oxides, dioxides and hydroxides 
form readily through gas-phase chemistry. What is surprising, and certainly not predicted 
by the TE models is the presence of carbon-bearing molecules since all available carbon is 
expected to be locked up in CO.  HCN, CS and CO$_2$, hoever, have relatively large
abundances in the inner CSE.  Gobrecht et al.\cite{gob16} show that the chemistry, while
complex, occurs on very fast time-scales. Thus CS forms in hot gas via:

\begin{eqnarray}
S + H_2 & \longrightarrow & SH + H \\
C + SH & \longrightarrow & CS + H \\
CO + SH & \longrightarrow & OCS + H \\
OCS + H & \longrightarrow & CS + OH 
\end{eqnarray}
while CN and HCN form via:

\begin{eqnarray}
N + CO & \longrightarrow & CN + O \\
N + CS & \longrightarrow & CN + S 
\end{eqnarray}
followed by
\begin{equation}
CN + H_2 \longrightarrow  HCN + H
\end{equation}

Many of these reactions produce atomic hydrogen and the reverse reactions can be significant 
particularly when the abundance of H atoms is high. This occurs in the zone where
dust precursors form since their formation converts some H$_2$ to H, as discussed below.
Gobrecht et al. show that the gas-phase abundances calculated in their pulsational shock
model at 6--8 R$_{\star}$ match to within an order of magnitude those observed in the inner
CSE, including HCN and CS.  Some molecules, not surprisingly, do not fit as well, for
example, SO and SO$_2$. Danilovich et al. \cite{dan16} have recently observed many
transitions of these two molecules in R Dor and supplemented these with HIFI and other
observations toward another four O-rich AGB stars. Using a detailed radiative transfer model
they have constrained the distributions of these two species, both in abundance and
in radial extent. In all five stars they find that SO$_2$ peaks on the stellar position 
whereas SO has a shell-like structure in IK Tau and R Cas, with a peak fractional abundance
of $\sim$ 10$^{-6}$ at 1.3 $\times 10^{16}$ cm.

When the peak position of the SO abundance is plotted against the wind density, $\propto 
\dot{M}/v_{exp}$, the results for R Cas, TX Cam and IK Tau follow a power-law
dependence consistent with a circumstellar chemistry that is dominated by photodissociation
in the outer envelope. SO$_2$, on the other hand, appears to constrained to the inner envelope
where it should be formed in high-temperature chemistry. Although the latter fact is broadly consistent 
with the models of Gobrecht et al., the observed SO$_2$ abundances in R Dor and W Hya are an
order of magnitude larger than predicted. Indeed all models fail to predict the very large 
abundances of SO and SO$_2$ which, in total, approach the elemental sulphur abundance. Model
calculations generally predict sulphur to be in atomic form.  A failure to agree in every 
respect with the observations should, however, be qualified by noting that an accurate description
of the chemical kinetics occurring at high density and temperature is very difficult, 
especially in the case of O-rich AGB stars since inorganic chemistry is not so well studied in the laboratory.
In addition, the balance between forward and reverse reactions, such as several of those above,
is controlled by the H:H$_2$ abundance ratio which is not well determined either
observationally or theoretically in the inner envelope.

In addition to the synthesis of molecules such as CO, H$_2$O, PN and HCl, which do not 
participate in dust formation, many other species are produced, several of which are likely to be
intimately connected with the process of cluster formation and grain growth.  Goumans and
Bromley \cite{gou12} discussed the detailed energetics of the formation of the dimers of
enstatite (MgSiO$_3$)$_2$ and fosterite (Mg$_2$SiO$_4$)$_2$ from an initial gas of
SiO, Mg, H$_2$ and H$_2$O at 1000 K. Although the initial dimerisation of SiO is an endoergic
process, its equilibrium constant is $5.5 \times 10^{-4}$ at 1000 K, giving rise to a low 
abundance of Si$_2$O$_2$. Subsequent reactions with H$_2$O, which result in O-atom addition, 
followed by addition reactions with Mg are exothermic and can rapidly build dimers of both 
enstatite and forsterite. Since reactions of H$_2$O and Mg are exoergic for reactions with the 
dimers and larger clusters, silicate dust grains will form. Gobrecht et al \cite{gob16} 
find dimer fractional abundances of 10$^{-11}$ for enstatite and $5 \times 10^{-8}$ for forsterite
at 3.5 R$_{\star}$. Dimer formation becomes very efficient outside 3 R$_{\star}$ with growth
of silicate grains occurring between 3 and 6 R$_{\star}$. The authors follow the diffusion 
and coagulation of these particles to determine the grain size distribution as particles 
propagate outwards from 3.5 to 10 R$_{\star}$.  Their results show that, in most cases, gas-phase 
abundances agree well with those determined for the inner wind of IK Tau, that forsterite 
grains are much more abundant than enstatite and metal oxides such as MgO and SiO, and
that the overall dust-to-gas mass ratio is $\sim$ (1--6) $\times$ 10$^{-3}$, in reasonable 
agreement with observations.

Dust grains grow and their size distribution evolves over a number of pulsations as the gas 
is lifted slowly away from the photosphere.  Due to their high binding energy, clusters of 
alumina form readily in the hottest gas near the photosphere in O-rich stars. For a 
radial drift velocity of 0.5 km s$^{-1}$, it takes 12 pulsations for the gas in IK Tau to 
move between 1 and 2 R$_{\star}$. Gobrecht et al. find that the size distribution of alumina 
favours larger particles and that growth of alumina grains stops beyond 2 R$_{\star}$ as
all available aluminium is tied up in dust at that point.  Such grains make only a minor
contribution to the overall dust-to-gas mass ratio. 

Silicates, on the other hand, form at 
larger radii, out to about 10 R$_{\star}$, since formation of the underlying dimer population
cannot occur at high temperatures close to the star.  The drift velocity in the silicate
dust zone is larger, perhaps 1.5 km s$^{-1}$, than that in the alumina dust zone since 
radiation pressure on the alumina grains begins to drive the mass loss. In this case, it
takes about 10 pulsations to move material from 3.5 to 6 R$_{\star}$ where the majority
of silicate grains form. As these grains form further from the star where densities are lower, 
silicates tend to have a smaller size distribution that alumina grains. They do, however, 
because of the abundance of silicon, magnesium and oxygen, contribute more to the
dust mass. At 10 R$_{\star}$, the dust-to-gas mass ratio is $\sim 2 \times 10^{-3}$, similar to 
those observed in O-rich AGB stars, with $\sim 22\% $ of elemental silicon contained in 
the dust \cite{gob16}.

The outer CSE chemistry of O-rich AGB stars is dominated by effects produced by 
irradiation of the outer envelope by interstellar FUV photons, that is, outer CSEs are examples of
photon-dominated regions (PDRs). Li et al. \cite{li16} have presented a detailed model
of the outer CSE chemistry of O-rich CSEs including, for the first time, shielding of
N$_2$ in addition to the usual self- and mutual-shielding of H$_2$ and CO. The authors 
assume an extensive list of parent species, some 18, with initial conditions derived from
either observation or from the shock-induced abundances calculated at 6 R$_{\star}$ by
Gobrecht et al. \cite{gob16}. They determine the chemistry of some 467 species using the 
latest release of the The UMIST Database for Astrochemistry \cite{mce13}. Li et al. calculate 
radial abundances and column densities for mass-loss rates between 10$^{-8}$ and 10$^{-4}$
M$_{\odot}$ yr$^{-1}$ and expansion velocities of 10-40 km s$^{-1}$ and make specific
comparison with the observed abundances in IK Tau ($\dot{M}$ = $4.5 \times 10^{-6}$ 
M$_{\odot}$ yr$^{-1}$, $v_{exp}$ = 24 km s$^{-1}$).

Li et al. were able to include a detailed consideration of N$_2$ photodissociation due
to over 25 years of laboratory and theoretical studies \cite{aje89, ndo08, hea11}.
The rate coefficient is determined primarily by the overlap of the N$_2$ absorption bands
with those of H$_2$ (mutual shielding) in the 912--1000 ${\rm \AA}$ wavelength range together with
self-shielding of N$_2$. The overall shielding is thus a complex function of gas temperature
and column density \cite{li13}. For parameters appropriate to IK Tau, for example, 
the radius at which the fractional abundance of atomic nitrogen, produced by the 
photodissociation of N$_2$, reaches 10$^{-5}$, increases from 1 to 6 $\times$ 10$^{16}$ cm;
at a fractional abundance of 10$^{-4}$ the increase is from 2.8 to 18 $\times$ 10$^{16}$ cm.
Thus the atomic N abundance increases over an appreciable volume of the outer envelope. It
has, however, only a limited impact on molecular abundances of species other than N$_2$, 
in part because N is a fairly unreactive element at low temperatures and because the gas 
number density is low, $\sim 2 \times 10^4 (r/10^{16})^{-2}$ cm$^{-3}$, and hence 
collision times are long. One molecule that shows a large difference when the N$_2$
shielding is modelled correctly is NO, produced by the N + OH reaction, with its peak 
fractional abundance decreasing by over an order of magnitude from $\sim 10^{-6}$ to
6 $\times 10^{-8}$ at a radius of 2.5 $\times 10^{16}$ cm.  For parameters appropriate
to IK Tau, the increased abundance of N$_2$
in the outer CSE leads to an increased N$_2$H$^+$ abundance due to the proton transfer 
reaction, H$_3^+$ + N$_2$ $\longrightarrow$ N$_2$H$^+$ + H$_2$, at $r < 10^{17}$ cm, 
and, at larger radii, the reaction He$^+$ + N$_2$ $\longrightarrow$ N$_2^+$ + He followed
by N$_2^+$ + H$_2$ $\longrightarrow$ N$_2$H$^+$ + H.  The abundance of N$_2$H$^+$ is
directly correlated with the initial (unknown) abundance adopted for N$_2$ but its use as 
a tracer of N$_2$ is limited since its predicted column density is low, only 
3.4 $\times$ 10$^{10}$ cm$^{-2}$.

\subsection{C-rich late-type stars} \label{crich}
Carbon-rich AGB star envelopes experience the same physical processes as those around O-rich stars 
but their molecular content is very different, in both composition and complexity, primarily 
due to the reactive nature and unique bonding properties of the carbon atom.

Cherchneff \cite{che12} has produced the most detailed model for the non-equilibrium chemistry 
of the inner dust formation region of IRC+10216, by far the most
well observed carbon-rich AGB star. This star has a mass-loss rate of 1.5 $\times$ 10$^{-5}$
M$_{\odot}$ yr$^{-1}$, a terminal expansion velocity of 14.5 km s$^{-1}$ and is known to
contain at least 80 molecules in its CSE. The vast majority of these molecules are hydrocarbons,
well understood because of the high abundance of carbon relative to oxygen in this star. A 
surprise discovery, however, was the presence of cold water \cite{mel01}, OH \cite{for03} 
and H$_2$CO \cite{for04}. A number of explanations were put forward including the evaporation 
of icy bodies within the CSE \cite{for01} and the formation of water on metallic grains 
\cite{wil04} but all mechanisms had their problems. Subsequent to these observations, the 
{\it Herschel} satellite was used to survey water in a number of C-rich AGB stars and it was 
found that warm water was present in many \cite{dec10b, neu10}, indicating that abundant 
water was present close to the dust-forming regions in these stars.  In this scenario, 
alternative formation mechanisms become possible, most importantly, shock chemistry following
stellar pulsations \cite{che12}, similar to that in O-rich stars, and photon-driven chemistry 
following deep penetration of interstellar photons through a clumpy envelope \cite{agu10}.

More recently, Lombaert et al. \cite{lom16} present {\it Herschel} observations of H$_2$O 
toward 18 C-rich AGB stars to look for correlations between abundances, dynamics and physical
conditions. They find warm H$_2$O emission from all stars and conclude that water is located close to 
or inside the wind acceleration zone, i.e., the dust formation  zone since the wind is 
driven by momentum transfer from the dust to the gas \cite{kwo75}. Detailed excitation and radiative
transfer calculations indicate that the fractional abundance of water lies in the range
10$^{-6}$--10$^{-4}$, at maximum some two to three orders of magnitude larger than 
predicted by the UV photodissociation \cite{agu10} or shock chemistry \cite{che12} models.
In the latter model, a fraction of parent CO is collisionally destroyed in the immediate 
post-shock gas and the O atoms released take part in fast neutral reactions that either
reform CO or form oxides, most importantly H$_2$O and SiO. One should note that the shock 
model does predict a high fractional abundance, 10$^{-4}$, of H$_2$O inside the dust
formation zone at less than 2.5 R$_{\star}$ for the specific case of IRC+10216. In addition,
the large abundance variations, some six orders of magnitude, predicted within a pulsational phase
at these small radii should lead to variable emission in the high-energy water transitions.

Furthermore, these non-equilibrium shock models reproduce the abundances of several other species,
including NaCl, AlCl and KCl, to within an order of magnitude, remarkably well given the uncertainties
in many of the rate coefficients. Cherchneff \cite{che12} has also calculated the
abundance of simple hydrocarbons up to benzene, C$_6$H$_6$, which is known to be
necessary for the production of polycyclic aromatic hydrocarbons (PAHs) and, perhaps
more generally, for the formation of carbonaceous dust grains in C-rich AGB stars. The models
find that a large fractional abundance, $\sim$ 10$^{-6}$, of benzene forms late in the
pulsation, at phases greater than 0.8, when the gas is cool and the abundances of H$_2$O
and OH are low, less than 10$^{-6}$, since both species oxidise benzene and prevent the 
growth of larger PAH-like molecules. Her calculations, under the assumption that all
C$_6$H$_6$ is converted to coronene, C$_{24}$H$_{12}$, through reactions involving acetylene,
C$_2$H$_2$, and that the total mass of coronene ends up in dust, gives reasonable agreement with
the dust-to-gas mass ratio observed in IRC+10216.

The UV photodissociation model \cite{agu10} produces H$_2$O with a fractional abundance of (2--10)
$\times$ 10$^{-7}$ at 2--10 R$_{\star}$, much lower than observed for high mass-loss rate stars and
also depends critically on a significant degree of clumping and/or scattering of UV photons
to allow a few percent of the interstellar UV flux to reach radii less than 10$^{15}$ cm. At
this radius a spherically symmetric uniform outflow, with a mass-loss rate equivalent to that of IRC+10216, 
would have a radial UV extinction of more than 50 magnitudes.  The challenge for both the shock and 
the UV models is that in order to produce very high abundances of water, O atoms must be 
liberated efficiently from CO and processed by the chemistry away from CO to H$_2$O. For the 
UV model, the main isotopologue of CO, $^{12}$C$^{16}$O, has a very small photodissociation rate since it
self-shields efficiently and is mutually shielded by H$_2$, the same is likely true also for $^{13}$C$^{16}$O.
As a result water should be enhanced in $^{17}$O and $^{18}$O but this does not
seem to be the case \cite{neu13}. We note that isotope effects are not expected in the shock
model since CO is destroyed collisionally and not radiatively. Photodissociation of SiO 
may provide O atoms but, because of the cosmic abundance of silicon, cannot account for 
water fractional abundances much greater than 10$^{-5}$.

Observations at high spatial resolution with {\it ALMA} are now providing a remarkable
view of the inner envelope of IRC+10216, particularly those molecules that appear to be 
related to dust formation.  {\it ALMA} has been used to observe SiS, SiO and SiC$_2$ 
\cite{vel15} with different radial distributions pointing to different formation mechanisms.
SiS emission comes from a small region, radius $\sim$ 1" centred on the star, with SiO also 
peaking there but with a more extended distribution, $\sim$ 3--3.5" in radius. SiC$_2$, on the other hand, 
shows both a central peak but also a ring of emission with radius around 10", or 2 $\times$ 10$^{16}$ cm, 
consistent with a photochemical origin in the outer envelope. Of these molecules, SiS
is the most abundant, with the total abundance of the three molecules accounting for
a significant fraction of elemental silicon.  Recently, some 112 rotational detections of 
SiCSi were detected, the first disilicon molecule discovered in space \cite{cer15} and
a molecule predicted to be abundant in TE calculations.

One molecule detected in the inner CSE but not expected from TE or shock chemistry, is
CH$_3$CN which has a hollow shell distribution with inner and outer radii of 1" and 2",
respectively \cite{agu15}. In interstellar clouds CH$_3$CN is formed by the fast
radiative association of CH$_3^+$ with HCN followed by dissociative recombination with
electrons, with likely a minor contribution from ice chemistry in regions where that is
important.  The very large abundance of HCN in the inner envelope clearly helps produce
CH$_3$CN but if the ultimate source of the ionisation in the inner envelope is cosmic
ray protons, as it is in the dense interstellar clouds in which CH$_3$CN is observed, 
then the abundance of CH$_3^+$ is likely to be vanishingly low for two reasons. One
is that the ionisation rate cannot be larger than 10$^{-17}$ s$^{-1}$, a constraint
imposed by the very low abundance of HCO$^+$ detected in IRC+10216 \cite{pul11}. The
second is that the ionisation fraction generally decreases as $1/n$ in dense gas so
that the formation of ions is less efficient in the inner envelope than further out
\cite{agu15}.  An alternative explanation has been considered \cite{agu15}, namely that 
a few percent of interstellar FUV photons incident on the external envelope penetrate down to 
or close to the photosphere. In this case the CH$_3$CN abundance increases by about
two orders of magnitude inside 8", although the distribution is centrally peaked on the 
star rather than distributed in a hollow shell \cite{agu15}.

The spectacular and complex hydrocarbon chemistry of the outer CSE in IRC+10216 has been explored
by a number of authors (\cite{hug84, nej87, che93b, dot98, mil00, cor09, li14}). Here, the
most important species are parents such as C$_2$H$_2$ and HCN whose photodissociation and photoionisation
provides a rich reactive soup of radicals, atoms and ions that rapidly build long-chain
hydrocarbons. Photodissociation of parent molecules gives rise to the ring distributions
seen in daughter species such as C$_2$H and CN. These radicals react with other radicals 
as well as parents to build complexity, e.g., the reactions
\begin{eqnarray}
C_2H + C_2H_2 & \longrightarrow & C_4H_2 + H\\
C_2H_2^+ + C_2H_2 & \longrightarrow & C_4H_3^+ + H \\
C_2H_2^+ + C_2H_2 & \longrightarrow & C_4H_2^+ + H_2 \\
CN + C_2H_2 & \longrightarrow & HC_3N + H
\end{eqnarray}
rapidly form abundant C$_4$-bearing hydrocarbon neutrals such as C$_4$H and H$_2$CCCC, and
cyanoacetylene, HC$_3$N.

In a model calculation containing molecules with up to 
23 carbon atoms, Millar et al. \cite{mil00} show that simple synthetic pathways give rise 
to efficient growth in molecular size and to ring distributions as observed. For specific
classes of molecules, such as the cyanopolyynes or alkenes, they find that 
peak fractional abundances and column densities typically fall by a factor of 2--3 as the
number of carbon atoms increase. For a constant mass-loss rate and expansion velocity, the
increased time to make larger molecules from smaller species results in radial distributions
in which the position of the peak abundance generally increases as molecular size increases.
Thus, for example, the peak fractional abundance of C$_2$H is reached at 4.0 $\times$ 10$^{16}$ cm
while that for C$_7$H occurs at 7.1 $\times$ 10$^{16}$ cm. This type of behaviour is not always
seen in the observations \cite{gue99} indicating that either the chemistry is more complex, 
occurring in parallel rather than sequentially, or that molecules are being produced by processes 
involving grains.  The molecular shells of HC$_3$N and HC$_5$N are found to be clumpy, co-spatial 
and with a distribution that closely matches that of dust shells and arcs in the outer CSE of 
IRC+10216 \cite{din08}. These shells are also seen in CO emission out to a radius of around 180" where CO is photodissociated
\cite{cer15b}. Surprisingly, these shells are not centred on the star itself suggesting that 
these periods of enhanced mass loss are induced at periastron by a companion star. 
A more physically realistic model of this envelope, taking into account
the presence of enhanced density shells in both gas and dust was produced by Cordiner and Millar \cite{cor09}. 
They based their idealised model on observations and added eight shells, each 2" thick, 
with an overdensity of 5 compared to the normal $1/r^2$ distribution, and an intershell spacing of 12".
Assuming a distance of 130 pc to IRC+10216 and the observed expansion velocity of 14.5 km s$^{-1}$, 
this corresponds to an enhanced mass-loss rate occurring for 90 years every 530 years.

The inclusion of shells, not surprisingly, causes the radial distributions of molecules to be
better aligned to one another and to the dust. Chemistry is enhanced within the shells since
the reaction time goes as $n^{-2}$. In addition, the shells provide additional extinction to the
penetration of external UV photons and move the inner edge of the molecular ring distributions outward.
For example, the peak fractional abundances of HC$_3$N and HC$_5$N move from 8" to 15" when shells are included.
Cordiner and Millar \cite{cor09} find that the shell at 15" dominates the emission characteristics of a number
of molecules, that is, the model predicts rings of co-spatial emission from C$_2$H, C$_4$H
and C$_6$H and from HC$_3$N and HC$_5$N, as observed.

One of the major successes of the photochemical modelling of IRC+10216 has been the 
prediction and subsequent detection of several large anions in the outer CSE. In the 
past ten years or so, laboratory measurements of the microwave spectra of anions has led 
to the identification of C$_4$H$^-$, C$_6$H$^-$, C$_8$H$^-$, CN$^-$, C$_3$N$^-$ and
C$_5$N$^-$ in IRC+10216.  Such anions were predicted with abundances that could be a 
significant fraction of their neutral analogues, for example the C$_8$H$^-$/C$_8$H column density 
ratio was predicted to be 0.25 \cite{mil00} and observed to be 0.26 \cite{rem07}.  These
anions are formed predominantly through the radiative attachment of electrons to neutral 
hydrocarbons which possess large electron affinities. For molecules with five or more carbon 
atoms, the attachment occurs on almost every collision.  The abundance of anions in 
IRC+10216 is so large that there are regions in the envelope in which the anion abundance
exceeds that of free electrons. They are also very reactive and play an important role in 
the synthesis of even larger hydrocarbon species \cite{cor09}.

The most recent release of the UMIST Database for Astrochemistry (www.udfa.net) now
contains over 20 anions involved in some 1300 gas-phase reactions. The full UDfA database,
some 6173 reactions among 467 species, was used to study chemistry in a model of IRC+10216
assuming a constant mass-loss rate \cite{mce13}. Some 31 out of 47 of the `daughter' species
were found to have column densities that agreed to within an order of magnitude of those
observed, indicating that we understand in broad terms the nature of the chemistry in 
carbon-rich circumstellar envelopes in AGB stars.  This sort of agreement with observation 
implies that we have a fairly complete knowledge of the gas-phase chemical kinetics that
occurs in the outer envelope of IRC+10216. Remaining uncertainties are linked either to 
unknown rate coefficents, primarily photodissociation rates, and reactions involving large
hydrocarbon ions and neutrals, or to uncertain or unknown abundances of parent molecules or
to physical structures within the CSE.  The situation in in the inner CSE, roughly defined 
here as interior to 10$^{16}$ cm, densities greater than 10$^{5-6}$ cm$^{-3}$ and temperatures 
greater than 100K is still open to significant improvement in understanding. We have already 
mentioned some areas in relation to both shock chemistry and FUV-dominated chemistry near
the photosphere. The role of dust grains, once formed and driven outward by radiation pressure,
is unexplored and there is observational evidence that they can act both as sinks and
sources of gas-phase molecules.

\section{Discussion}
\label{disc}

\begin{figure}[th]
\begin{minipage}{16pc}
\includegraphics[width=16pc,angle=-90,scale=0.8]{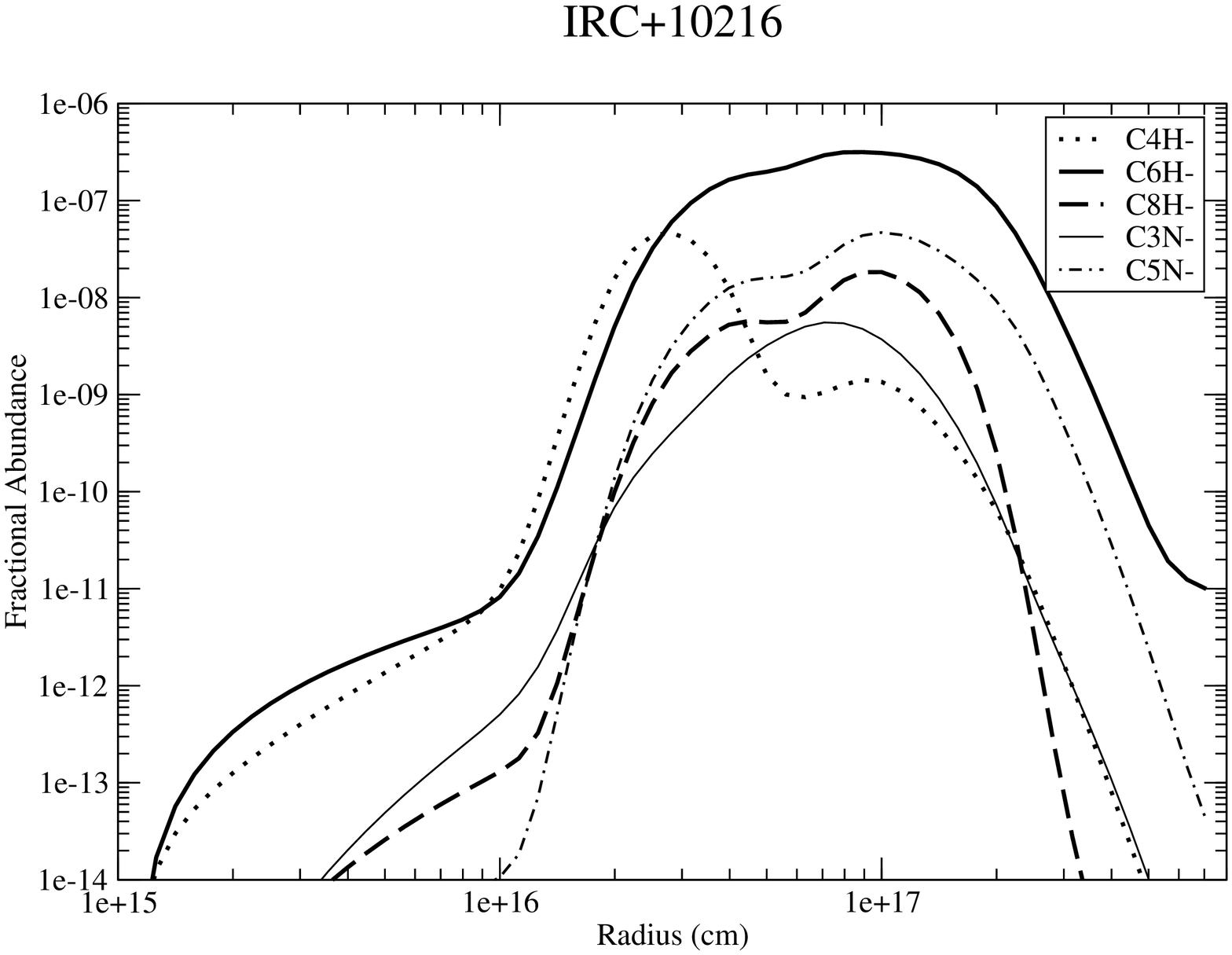}
\caption{Fractional abundance of anions versus radius for the standard spherically symmetric outflow.}
\label{fig:1}
\end{minipage}\hspace{2pc}%
\begin{minipage}{16pc}
\includegraphics[width=16pc,angle=-90,scale=0.8]{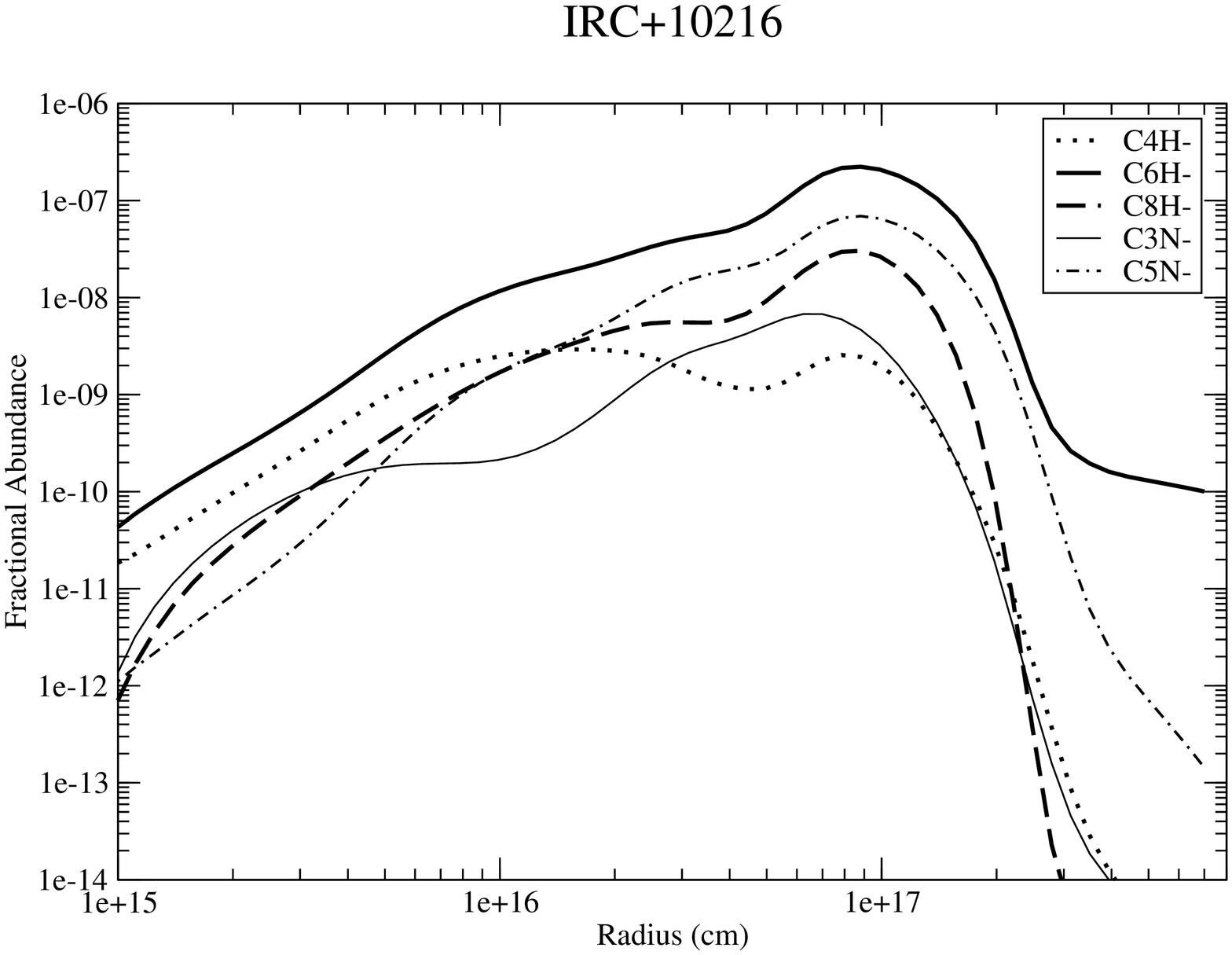}
\caption{As figure 1 but with 2.5\% of interstellar photons able to penetrate deep into the inner CSE.}
\label{fig:2}
\end{minipage} 
\end{figure}

As outlined above, the discovery of H$_2$O in C-rich CSEs, and more recently CS and HCN
in O-rich CSEs, indicates that there are processes that perturb the TE chemistry that is 
expected to dominate at and close to the stellar photosphere.  Shock chemistry induced by
stellar pulsations is clearly important as may be the detailed chemistry associated with 
dust formation and growth. Despite the advances that have been made recently, this still 
remains a poorly understood area with a lack of critical experimental data appropriate
to the densities and temperatures found in the dust formation zone.  The penetration of FUV 
photons deep into the CSE is another possible mechanism. The models discussed above 
require that a significant fraction, some 2.5\% \cite{agu10,agu15}, of interstellar photons
need to avoid around 30 magnitudes of FUV extinction that lie between 1" (2 $\times$ 10$^{15}$
cm) and the edge of the CSE.  If this occurs then it has profound effects on the 
composition of the gas in the inner regions of both C-rich and O-rich CSEs \cite{agu10,agu15}.
Figure ~\ref{fig:1} shows the radial distributions of the fractional abundances of some hydrocarbon 
anions calculated under standard conditions, i.e. a spherically symmetric outflow at constant 
mass-loss rate, with interstellar UV photons incident on the outer CSE. I have, in 
figure ~\ref{fig:2}, adopted this model to allow the same percentage of FUV 
photons to reach 10$^{15}$ cm unaffected by dust extinction \cite{mil00,cor09}, a useful exercise since I 
calculate the FUV radiative transfer in a different way to that of Ag{\'u}ndez and collaborators.
It can be seen that the radial distributions of the anions, a representative class of the
hydrocarbons, show significant differences particularly inside (3--4) $\times$ 10$^{16}$ cm, 
when FUV photons are allowed to penetrate. The fractional abundances typically increase by 
about two orders of magnitude inside 10$^{16}$ cm although the change in column density is less pronounced,
typically 2--3 for these species and generally less than  a factor of five for most others.

Although FUV photons do tend to increase the fractional abundances down to a few 
10$^{15}$ cm, the distributions do not show the sharp inner boundaries evident in some 
emission maps \cite{din08,agu15}. Could these sharp inner edges be an indicator that stellar 
photons are responsible? To date, the role of such photons has been ignored on two grounds.
The first is that IRC+10216 is too cool (T$_{eff}$ = 2330 K) to produce UV photons, the second 
that the dust-forming zone will provide a large amount of extinction, several hundred
magnitudes at UV wavelengths, to photons generated by the star. If, however, the mass 
loss and dust formation processes
themselves produce the clumpy structures that are inferred beyond 1" then it remains a 
possibility that some stellar photons do leak out to 50 R$_{\star}$.

The fact that the star is cool will indeed imply that the flux of photons at wavelengths 
less than 2000 ${\rm \AA}$ is small. There are, however, a number of molecules which have 
relatively small bond energies and which can be destroyed by photons at longer wavelengths.
Examples include the hydrocarbon anions, C$_{2{\rm n}}$H$^-$, n = 1--4, which have 
electron affinities (EA) that range from 3.02 eV (C$_2$H$^-$) to 3.96 eV (C$_8$H$^-$). The
photodetachment cross-sections of these anions may be calculated using the empirical
formula \cite{mil07}:
\begin{equation}
\sigma(\epsilon) = \sigma_{\infty}\left (1 - \frac{EA}{\epsilon} \right )^{1/2}
\end{equation}
where $\epsilon$ is the photon energy and $\sigma_{\infty}$ is the asymptotic cross-section
at large energies. Data for EA and $\sigma_{\infty}$ have been provided experimentally
\cite{bes11,kum13}. At a distance of 50 R$_{\star}$ and assuming no extinction due to dust,
the electron photodetachment rates vary from 10$^{-5}$ s$^{-1}$ (CN$^-$) to 1.98 $\times$
10$^{-6}$ s$^{-1}$ (C$_6$H$^-$), Other species that have large photodissociation rates include
CH, l-C$_3$H, C$_5$H and NaCl. Thus, if even a small fraction of these stellar photons 
penetrate the dust-formation region and beyond, they could have a significant effect on the 
radial distribution of some species.  Detailed calculations investigating this are underway.

\ack

I should like to that the organisers for inviting me to participate in this conference. Astrophysics
at QUB is supported by a grant from the STFC.

\section*{References}

\bibliography{millar_hk}

\end{document}